\documentclass[10pt,aps,prd,twocolumn,superscriptaddress,nofootinbib,floatfix]{revtex4-1}
\usepackage{amsmath,amssymb}
\usepackage{graphicx}
\usepackage{hyperref}
\usepackage{capt-of}
\AtBeginDocument{\hfuzz=1000pt\vfuzz=1000pt}

\def\re#1{\mathrm{Re}(#1)}
\def\im#1{\mathrm{Im}(#1)}

\def\Order#1{{\cal O}\left(#1\right)}

\begin{document}

\title{Quasinormal Modes of a Massive Scalar Field in 4D Einstein--Gauss--Bonnet Black Hole Spacetimes}
\author{Bekir Can L\"utf\"uo\u{g}lu}
\email{bekir.lutfuoglu@uhk.cz}
\affiliation{Department of Physics, Faculty of Science, University of Hradec Kr\'alov\'e, Rokitansk\'eho 62/26, 500 03 Hradec Kr\'alov\'e, Czech Republic}

\begin{abstract}
We analyze quasinormal modes, grey-body factors, and absorption cross-sections of a massive scalar field in four-dimensional Einstein--Gauss--Bonnet black-hole spacetimes within a stability-constrained coupling window. High-order WKB-Pad\'e spectra show that increasing field mass typically reduces damping and drives the system toward long-lived, quasi-resonant behavior. The scattering sector follows the same potential-barrier physics: larger effective barriers suppress transmission and low-frequency absorption, while the Gauss--Bonnet coupling has a comparatively mild impact over the stable range. These results provide a compact baseline for massive-field spectroscopy in higher-curvature black-hole backgrounds.
\end{abstract}

\maketitle

\section{Introduction}

Quasinormal modes (QNMs) are the characteristic damped oscillations of perturbed black holes. They are a direct link between theory and data because ringdown measurements constrain remnant parameters and test general relativity in the strong-field regime \cite{LIGOScientific:2016aoc,LIGOScientific:2017vwq,LIGOScientific:2020zkf,Babak:2017tow}. In modified-gravity settings, QNM frequencies and damping rates provide a sensitive diagnostic of how non-Einstein couplings deform black-hole dynamics \cite{Kokkotas:1999bd, Konoplya:2011qq, Berti:2009kk, Bolokhov:2025uxz}. This motivates precision QNM studies on backgrounds beyond Schwarzschild and Kerr.

Grey-body factors (GBFs) and absorption cross-sections are complementary parts of the same scattering problem. While QNMs probe the complex resonance spectrum, GBFs quantify transmission through the effective potential barrier and therefore control the frequency-dependent absorption/emission efficiency. Studying QNMs and GBFs together provides a more complete picture of wave propagation around black holes and helps determine whether observed trends are genuinely spectral or merely barrier-shape effects.

Higher-curvature corrections are a natural arena for such tests. They arise in effective descriptions of quantum gravity (for example, in low-energy string-inspired models) and can produce measurable strong-field deviations from general relativity. Among them, Einstein--Gauss--Bonnet (EGB)-type theories are especially useful because they provide a tractable one-parameter deformation with well-developed perturbation literature \cite{Konoplya:2017lhs,Churilova:2019sah,Blazquez-Salcedo:2020caw,Matyjasek:2020bzc,Konoplya:2020bxa,Chen:2015fuf,Staykov:2021dcj,Cano:2020cao,Konoplya:2017ymp,MoraisGraca:2016hmv,Kokkotas:2017zwt,Grozdanov:2016fkt,Blazquez-Salcedo:2022omw,Blazquez-Salcedo:2020rhf,Zinhailo:2018ska,Prasobh:2014zea,Kokkotas:2017ymc,Silva:2022srr,Konoplya:2022iyn,Zhidenko:2008fp,Sadeghi:2011zza,Blazquez-Salcedo:2024oek,Pierini:2022eim,Moura:2006pz,Daghigh:2006xg,Carson:2020ter,Chakrabarti:2006ei,Yoshida:2015vua,Zinhailo:2019rwd,Gonzalez:2018xrq,Abdalla:2005hu,Wagle:2021tam,Chakrabarti:2005cm,Mishra:2020gce,Cao:2023mai,Konoplya:2017zwo}. In this work we focus on the four-dimensional EGB black hole and explicitly restrict the Gauss--Bonnet coupling to a conservative stability window, since sufficiently large coupling can trigger gravitational eikonal instability.

Massive fields are important because the mass term can qualitatively reshape the spectrum and scattering observables. Depending on the background, increasing field mass may produce long-lived or quasi-resonant modes, modify late-time tails, and shift transmission/absorption thresholds \cite{Konoplya:2004wg,Gonzalez:2022upu,Zhidenko:2006rs,Aragon:2020teq,Ponglertsakul:2020ufm,Konoplya:2017tvu,Ohashi:2004wr,Burikham:2017gdm,Bolokhov:2023ruj,Zhang:2018jgj,Konoplya:2018qov,Koyama:2001ee, Koyama:2000hj, Gibbons:2008gg, Jing:2004zb, Konoplya:2007zx, Rogatko:2007zz, Moderski:2001tk, Lutfuoglu:2026fpx}. The massive term is also well motivated physically: effective masses can arise in brane-world scenarios \cite{Seahra:2004fg}, in magnetized black-hole environments \cite{Wu:2015fwa,Konoplya:2008hj}. In addition, ultralow-frequency gravitational-wave constraints \cite{NANOGrav:2023hvm} motivate mapping these mass-dependent effects in detail \cite{Konoplya:2023fmh}.

From the observational side, this program is timely. Ringdown analyses already play a central role in LIGO--Virgo--KAGRA tests of strong-field gravity \cite{KAGRA:2021vkt,KAGRA:2013rdx}. Future detectors strengthen this motivation: LISA will target high-SNR massive-black-hole ringdowns, while third-generation ground-based detectors (Einstein Telescope and Cosmic Explorer) are designed to significantly improve ringdown precision for stellar-mass remnants.

Quasinormal modes of massless fields in four-dimensional EGB gravity have been studied in detail in \cite{Konoplya:2020bxa}. In contrast, massive perturbations have so far been analyzed only in \cite{Becar:2025niq} for asymptotically de Sitter black holes, where lower-order WKB approximations were employed and analytic expressions were obtained in the limit of large field mass $\mu$. However, in the limit of zero cosmological constant these expressions are not valid.

In this paper, we compute QNMs, GBFs, and absorption cross-sections for a {\it massive} scalar field on the asymptotically flat 4D-EGB background using two complementary approaches: a frequency-domain high-order WKB analysis and a time-domain characteristic integration for cross-validation. We then interpret both spectral and scattering results in terms of the effective potential and assess their dependence on the scalar mass and Gauss--Bonnet coupling within the stability-admissible region.

The paper is organized as follows. In Sec.~\ref{sec2} we introduce the background geometry and derive the massive-scalar perturbation equation with its effective potential. Section~\ref{sec3} summarizes the frequency-domain WKB framework, and Sec.~\ref{sec4} presents the time-domain characteristic method. In Sec.~\ref{sec5} we define the stability-constrained parameter domain used throughout the analysis. Sections~\ref{sec6} and \ref{sec7} report, respectively, the QNM results and the grey-body/absorption results, including their physical interpretation. We conclude in Sec.~\ref{sec8} with a summary and outlook.

\section{Background Geometry and Scalar Perturbation Equation} \label{sec2}

We work within the four-dimensional EGB framework obtained through the regularized $D\to4$ limit of the Einstein--Lovelock action, following the construction discussed in \cite{Glavan:2019inb}. In this procedure, the Gauss--Bonnet coupling is rescaled as $\alpha/(D-4)$ before taking the limit, so that the Gauss--Bonnet invariant
\begin{equation}
\mathcal{G}=R^2-4R_{\mu\nu}R^{\mu\nu}+R_{\mu\nu\rho\sigma}R^{\mu\nu\rho\sigma},
\end{equation}
contributes to the effective equations of motion in four dimensions. At the same time, we keep in mind the well-defined covariant reinterpretation of this setup in terms of a scalar-tensor realization (see \cite{Aoki:2020lig,Aoki:2020iwm}), where the same static black-hole geometry is recovered without relying on an ill-defined standalone four-dimensional Gauss--Bonnet variation. In this sense, the metric employed here can be viewed as common to both descriptions, and our perturbative analysis is performed at the level of that shared black-hole background. For static spherical symmetry, the metric function has two branches, and throughout this work we adopt the minus branch because it is asymptotically flat and continuously reduces to the Schwarzschild form $f(r)\to 1-2M/r$ as $\alpha\to0$.

We consider a static, spherically symmetric line element
\begin{equation}
 ds^2=-f(r)dt^2+\frac{dr^2}{f(r)}+r^2\left(d\theta^2+\sin^2\theta\,d\phi^2\right),
\end{equation}
with the four-dimensional EGB lapse function (minus branch)
\begin{equation}
 f(r)=1+\frac{r^2}{2\alpha}\left(1-\sqrt{1+\frac{8\alpha M}{r^3}}\right),
\end{equation}
where $M$ is the mass parameter and $\alpha$ is the Gauss--Bonnet coupling.

A massive scalar test field $\Phi$ obeys the Klein--Gordon equation
\begin{equation}
 (\Box-\mu^2)\Phi=0,
\end{equation}
where $\mu$ is the scalar mass. Introducing the standard decomposition
\begin{equation}
 \Phi(t,r,\theta,\phi)=e^{-i\omega t}Y_{\ell m}(\theta,\phi)\frac{\psi(r)}{r},
\end{equation}
and the tortoise coordinate $r_*$ via
\begin{equation}
 \frac{dr_*}{dr}=\frac{1}{f(r)},
\end{equation}
we obtain the Schr\"odinger-type radial equation
\begin{equation}
 \frac{d^2\psi}{dr_*^2}+\bigl[\omega^2-V_\ell(r)\bigr]\psi=0,
\end{equation}
with effective potential
\begin{equation}
 V_\ell(r)=f(r)\left[\frac{\ell(\ell+1)}{r^2}+\frac{f'(r)}{r}+\mu^2\right].
\end{equation}

\begin{figure}
\includegraphics[width=\columnwidth]{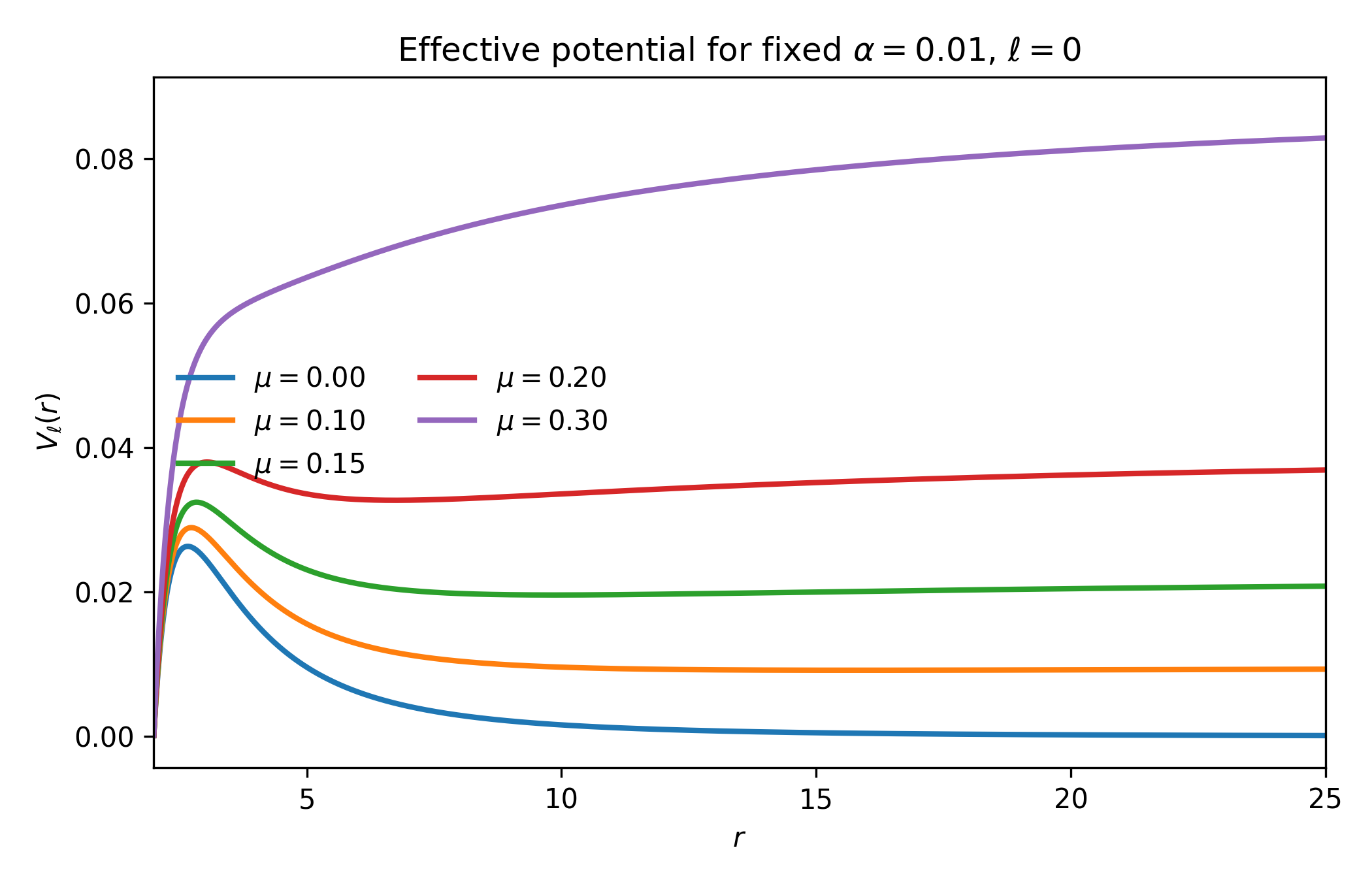}
\caption{Effective potential profiles $V_\ell(r)$ for fixed Gauss--Bonnet coupling $\alpha=0.01$ and $\ell=0$, with several scalar masses $\mu$. For small $\mu$, the potential has a clear barrier maximum; as $\mu$ increases, the barrier flattens and disappears (here around $\mu\approx0.2$), marking the onset of a no-peak regime.}
\label{fig:veff_mu_scan}
\end{figure}

\begin{figure}
\includegraphics[width=\columnwidth]{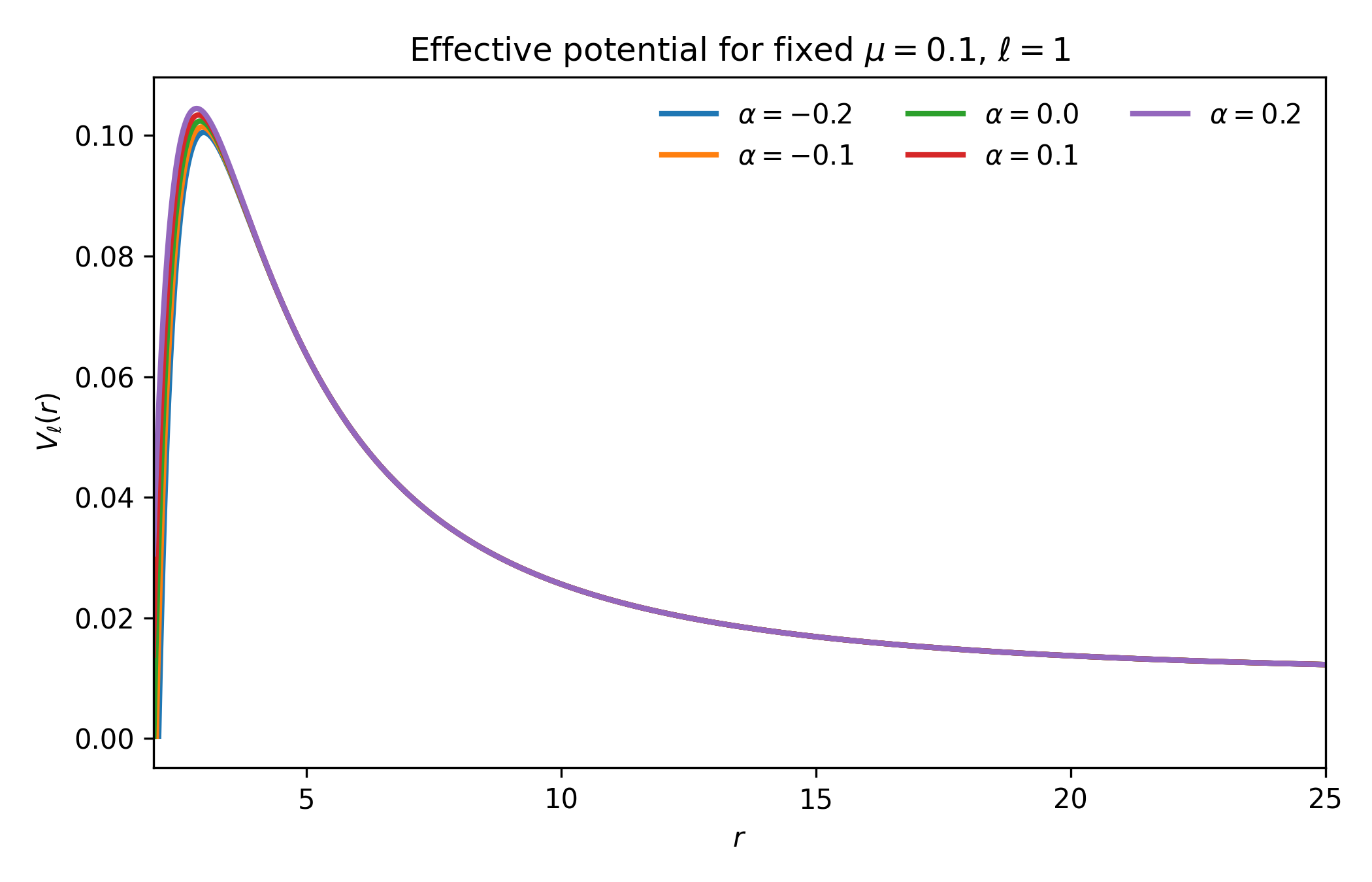}
\caption{Effective potential profiles $V_\ell(r)$ for fixed scalar mass $\mu=0.1$ and $\ell=1$, with different Gauss--Bonnet couplings $\alpha\in[-0.2,0.2]$. The overall potential shape changes only weakly across this range, illustrating the mild $\alpha$-dependence compared with the stronger $\mu$-dependence in Fig.~\ref{fig:veff_mu_scan}.}
\label{fig:veff_alpha_scan}
\end{figure}

For QNMs, boundary conditions are purely ingoing at the event horizon and purely outgoing at spatial infinity. The examples of effective potentials are shown in Figs.~\ref{fig:veff_mu_scan} and \ref{fig:veff_alpha_scan}. 

\section{Frequency-Domain Method: WKB Approximation} \label{sec3}

To compute complex frequencies $\omega=\omega_R-i\omega_I$ (with $\omega_I>0$ for damping), we use higher-order WKB expansion around the maximum of $V_\ell(r)$ \cite{Iyer:1986np,Konoplya:2003ii,Matyjasek:2017psv,Matyjasek:2019eeu}. Denoting by $V_0$ and $V_0''$ the value and second derivative (with respect to $r_*$) at the potential peak, the WKB condition is written as
\begin{equation}
 \frac{i\left(\omega^2-V_0\right)}{\sqrt{-2V_0''}}-\sum_{j=2}^{N}\Lambda_j=n+\frac{1}{2},
\end{equation}
where $n$ is overtone number and $\Lambda_j$ are higher-order correction terms.

In practical runs, one can evaluate both various higher order truncations as an internal convergence check, following standard QNM practice \cite{Stuchlik:2025ezz,Konoplya:2023moy,Arbelaez:2026eaz,Skvortsova:2024atk,Bolokhov:2023dxq,Kanti:2006ua,Lutfuoglu:2026xlo,Han:2026fpn,Malik:2024tuf,Konoplya:2025hgp,Bolokhov:2024bke,Skvortsova:2023zmj,Konoplya:2006ar,Lutfuoglu:2026gis,Konoplya:2002wt}. The method is typically most reliable for modes with $\ell>n$, and this criterion will be used in selecting benchmark points. For each $(\alpha,\mu,\ell,n)$ in the scan, the WKB output provides the first estimate of the QNM spectrum.

\section{Time-Domain Method: Characteristic Integration} \label{sec4}

To independently validate frequency-domain results, we evolve the perturbation in null coordinates
\begin{equation}
 u=t-r_*,\qquad v=t+r_*,
\end{equation}
for which the wave equation becomes
\begin{equation}
 4\,\partial_u\partial_v\psi+V(r)\psi=0.
\end{equation}

On a uniform null grid, we use the standard update formula
\begin{equation}
 \psi_N=\psi_W+\psi_E-\psi_S-\frac{\Delta^2}{8}V_S\left(\psi_W+\psi_E\right)+\mathcal{O}(\Delta^4),
\end{equation}
where $S,E,W,N$ are the south/east/west/north points of a grid cell.

The extraction protocol is: choose localized Gaussian initial data, evolve the waveform at fixed observer location $r_*$, isolate ringdown, and fit the late-time signal with a Prony or matrix-pencil routine to obtain complex frequencies. Agreement with WKB frequencies across overlapping reliability regions serves as a consistency test \cite{Dubinsky:2024gwo,Konoplya:2014lha,Skvortsova:2023zca,Stuchlik:2025mjj,Konoplya:2005et,Skvortsova:2025cah,Cuyubamba:2016cug,Konoplya:2013sba,Dubinsky:2024mwd,Lutfuoglu:2025bsf}.

\section{Stability-Constrained Parameter Domain} \label{sec5}

A central methodological point is that, in 4D-EGB black holes, sufficiently large Gauss--Bonnet coupling may trigger gravitational eikonal instability. Therefore, even though the present perturbation is a test scalar field, we restrict the scan to a gravitationally admissible coupling window.

For the same black-hole background, Konoplya and Zinhailo report the following gravitational-sector thresholds \cite{Konoplya:2020bxa}:
\begin{equation}
\alpha \lesssim 1.57M^2,\quad
\alpha \lesssim 0.6M^2,\quad
\alpha \gtrsim -16M^2.
\end{equation}
The strongest positive-coupling restriction comes from the scalar eikonal channel, while the negative-coupling bound is set by the vector channel. Their summary stability window is \cite{Konoplya:2020bxa}
\begin{equation}
-16M^2<\alpha\lesssim 0.6M^2,
\end{equation}
with instability setting in for sufficiently large positive coupling, i.e. $\alpha\gtrsim 0.6M^2$.

In this manuscript, we therefore set
\begin{equation}
\alpha\leq \alpha_{\max}^{\mathrm{grav}},\qquad \alpha_{\max}^{\mathrm{grav}}\leq 0.6M^2,
\end{equation}
and treat this as a conservative, literature-motivated cutoff. This prevents contaminating the massive-scalar QNM dataset with background points expected to be gravitationally unstable.

\section{Quasinormal modes} \label{sec6}

\begin{table}
\textbf{(a) $\alpha=-0.5$, $\ell=0$}\\
\resizebox{\columnwidth}{!}{%
\begin{tabular}{c c c c}
\hline
$\mu$ & WKB16 ($\tilde{m}=8$) & WKB14 ($\tilde{m}=7$) & difference \\
\hline
$0$ & $0.110554-0.104860 i$ & $0.110473-0.104522 i$ & $0.228\%$\\
$0.05$ & $0.111122-0.102849 i$ & $0.110997-0.102417 i$ & $0.297\%$\\
$0.1$ & $0.112943-0.096883 i$ & $0.115928-0.095736 i$ & $2.15\%$\\
$0.15$ & $0.115262-0.087814 i$ & $0.115504-0.087204 i$ & $0.452\%$\\
$0.2$ & $0.124930-0.068356 i$ & $0.114528-0.074718 i$ & $8.56\%$\\
\hline
\end{tabular}}

\vspace{0.1cm}
\textbf{(b) $\alpha=0.01$, $\ell=1$}\\
\resizebox{\columnwidth}{!}{%
\begin{tabular}{c c c c}
\hline
$\mu$ & WKB16 ($\tilde{m}=8$) & WKB14 ($\tilde{m}=7$) & difference \\
\hline
$0$ & $0.293057-0.097581 i$ & $0.293057-0.097581 i$ & $0\%$\\
$0.05$ & $0.294175-0.096910 i$ & $0.294175-0.096910 i$ & $0\%$\\
$0.1$ & $0.297533-0.094884 i$ & $0.297533-0.094884 i$ & $0\%$\\
$0.15$ & $0.303153-0.091455 i$ & $0.303153-0.091455 i$ & $0\%$\\
$0.2$ & $0.311065-0.086539 i$ & $0.311065-0.086539 i$ & $0\%$\\
$0.25$ & $0.321299-0.079998 i$ & $0.321299-0.079998 i$ & $0.\times 10^{-4}\%$\\
$0.3$ & $0.333868-0.071632 i$ & $0.333868-0.071632 i$ & $0.\times 10^{-4}\%$\\
$0.35$ & $0.348719-0.061165 i$ & $0.348721-0.061166 i$ & $0.00067\%$\\
$0.4$ & $0.365652-0.048323 i$ & $0.365671-0.048274 i$ & $0.0141\%$\\
$0.45$ & $0.384805-0.035835 i$ & $0.386401-0.036688 i$ & $0.468\%$\\
\hline
\end{tabular}}

\vspace{0.1cm}
\textbf{(c) $\alpha=0.01$, $\ell=2$}\\
\resizebox{\columnwidth}{!}{%
\begin{tabular}{c c c c}
\hline
$\mu$ & WKB16 ($\tilde{m}=8$) & WKB14 ($\tilde{m}=7$) & difference \\
\hline
$0$ & $0.483829-0.096683 i$ & $0.483829-0.096683 i$ & $0\%$\\
$0.05$ & $0.484618-0.096414 i$ & $0.484618-0.096414 i$ & $0\%$\\
$0.1$ & $0.486986-0.095602 i$ & $0.486986-0.095602 i$ & $0\%$\\
$0.15$ & $0.490942-0.094242 i$ & $0.490942-0.094242 i$ & $0\%$\\
$0.2$ & $0.496500-0.092324 i$ & $0.496500-0.092324 i$ & $0\%$\\
$0.25$ & $0.503678-0.089832 i$ & $0.503678-0.089832 i$ & $0\%$\\
$0.3$ & $0.512504-0.086743 i$ & $0.512504-0.086743 i$ & $0\%$\\
$0.35$ & $0.523010-0.083026 i$ & $0.523010-0.083026 i$ & $0\%$\\
$0.4$ & $0.535237-0.078640 i$ & $0.535237-0.078640 i$ & $0\%$\\
$0.45$ & $0.549232-0.073527 i$ & $0.549232-0.073527 i$ & $0\%$\\
$0.5$ & $0.565048-0.067609 i$ & $0.565048-0.067609 i$ & $0\%$\\
$0.55$ & $0.582741-0.060775 i$ & $0.582741-0.060775 i$ & $0\%$\\
$0.6$ & $0.602351-0.052873 i$ & $0.602352-0.052872 i$ & $0.00002\%$\\
$0.65$ & $0.623884-0.043705 i$ & $0.623883-0.043705 i$ & $0.00007\%$\\
$0.7$ & $0.647247-0.033027 i$ & $0.647208-0.033023 i$ & $0.00598\%$\\
\hline
\end{tabular}}
\caption{Fundamental scalar QNMs for $\alpha=0.01$ and $M=1$, computed with WKB-Pad\'e approximants $\tilde m=8$ (WKB16, i.e., 16th-order WKB) and $\tilde m=7$ (WKB14, i.e., 14th-order WKB). Panels (a)--(c) correspond to $\ell=0,1,2$. The last column reports the relative difference between the two approximants.}
\end{table}

\begin{table}
\textbf{(a) $\alpha=0.5$, $\ell=0$}\\
\resizebox{\columnwidth}{!}{%
\begin{tabular}{c c c c}
\hline
$\mu$ & WKB16 ($\tilde{m}=8$) & WKB14 ($\tilde{m}=7$) & difference \\
\hline
$0$ & $0.113954-0.099632 i$ & $0.113949-0.099656 i$ & $0.0165\%$\\
$0.05$ & $0.114587-0.097747 i$ & $0.114583-0.097786 i$ & $0.0261\%$\\
$0.1$ & $0.116302-0.093382 i$ & $0.116371-0.093320 i$ & $0.0620\%$\\
$0.15$ & $0.119777-0.083858 i$ & $0.120014-0.084366 i$ & $0.383\%$\\
$0.2$ & $0.120030-0.074233 i$ & $0.119866-0.074738 i$ & $0.376\%$\\
\hline
\end{tabular}}

\vspace{0.1cm}
\textbf{(b) $\alpha=0.5$, $\ell=1$}\\
\resizebox{\columnwidth}{!}{%
\begin{tabular}{c c c c}
\hline
$\mu$ & WKB16 ($\tilde{m}=8$) & WKB14 ($\tilde{m}=7$) & difference \\
\hline
$0$ & $0.299253-0.093368 i$ & $0.299254-0.093368 i$ & $0\%$\\
$0.05$ & $0.300328-0.092778 i$ & $0.300328-0.092778 i$ & $0\%$\\
$0.1$ & $0.303558-0.090992 i$ & $0.303558-0.090992 i$ & $0\%$\\
$0.15$ & $0.308964-0.087959 i$ & $0.308963-0.087959 i$ & $0.0001\%$\\
$0.2$ & $0.316575-0.083583 i$ & $0.316574-0.083583 i$ & $0.0002\%$\\
$0.25$ & $0.326423-0.077718 i$ & $0.326423-0.077718 i$ & $0.0001\%$\\
$0.3$ & $0.338524-0.070142 i$ & $0.338524-0.070142 i$ & $0.00003\%$\\
$0.35$ & $0.352825-0.060559 i$ & $0.352825-0.060557 i$ & $0.00037\%$\\
$0.4$ & $0.369139-0.048645 i$ & $0.369134-0.048662 i$ & $0.00472\%$\\
$0.45$ & $0.387259-0.035517 i$ & $0.387739-0.036159 i$ & $0.206\%$\\
\hline
\end{tabular}}

\vspace{0.1cm}
\textbf{(c) $\alpha=0.5$, $\ell=2$}\\
\resizebox{\columnwidth}{!}{%
\begin{tabular}{c c c c}
\hline
$\mu$ & WKB16 ($\tilde{m}=8$) & WKB14 ($\tilde{m}=7$) & difference \\
\hline
$0$ & $0.493505-0.092661 i$ & $0.493505-0.092661 i$ & $0\%$\\
$0.05$ & $0.494253-0.092425 i$ & $0.494253-0.092425 i$ & $0\%$\\
$0.1$ & $0.496501-0.091715 i$ & $0.496501-0.091715 i$ & $0\%$\\
$0.15$ & $0.500257-0.090523 i$ & $0.500257-0.090523 i$ & $0\%$\\
$0.2$ & $0.505536-0.088836 i$ & $0.505536-0.088836 i$ & $0\%$\\
$0.25$ & $0.512359-0.086637 i$ & $0.512359-0.086637 i$ & $0\%$\\
$0.3$ & $0.520755-0.083898 i$ & $0.520755-0.083898 i$ & $0\%$\\
$0.35$ & $0.530758-0.080584 i$ & $0.530759-0.080584 i$ & $0\%$\\
$0.4$ & $0.542414-0.076649 i$ & $0.542414-0.076649 i$ & $0\%$\\
$0.45$ & $0.555773-0.072028 i$ & $0.555773-0.072028 i$ & $0\%$\\
$0.5$ & $0.570894-0.066635 i$ & $0.570894-0.066635 i$ & $0\%$\\
$0.55$ & $0.587839-0.060349 i$ & $0.587839-0.060349 i$ & $0.00002\%$\\
$0.6$ & $0.606658-0.053008 i$ & $0.606658-0.053008 i$ & $0.00007\%$\\
$0.65$ & $0.627369-0.044402 i$ & $0.627370-0.044404 i$ & $0.00042\%$\\
$0.7$ & $0.649892-0.034358 i$ & $0.650007-0.034296 i$ & $0.0200\%$\\
\hline
\end{tabular}}
\caption{Fundamental scalar QNMs for $\alpha=0.5$ and $M=1$, computed with WKB-Pad\'e approximants $\tilde m=8$ (WKB16, i.e., 16th-order WKB) and $\tilde m=7$ (WKB14, i.e., 14th-order WKB). Panels (a)--(c) correspond to $\ell=0,1,2$. The last column reports the relative difference between the two approximants.}
\end{table}

\begin{figure}
\includegraphics[width=\columnwidth]{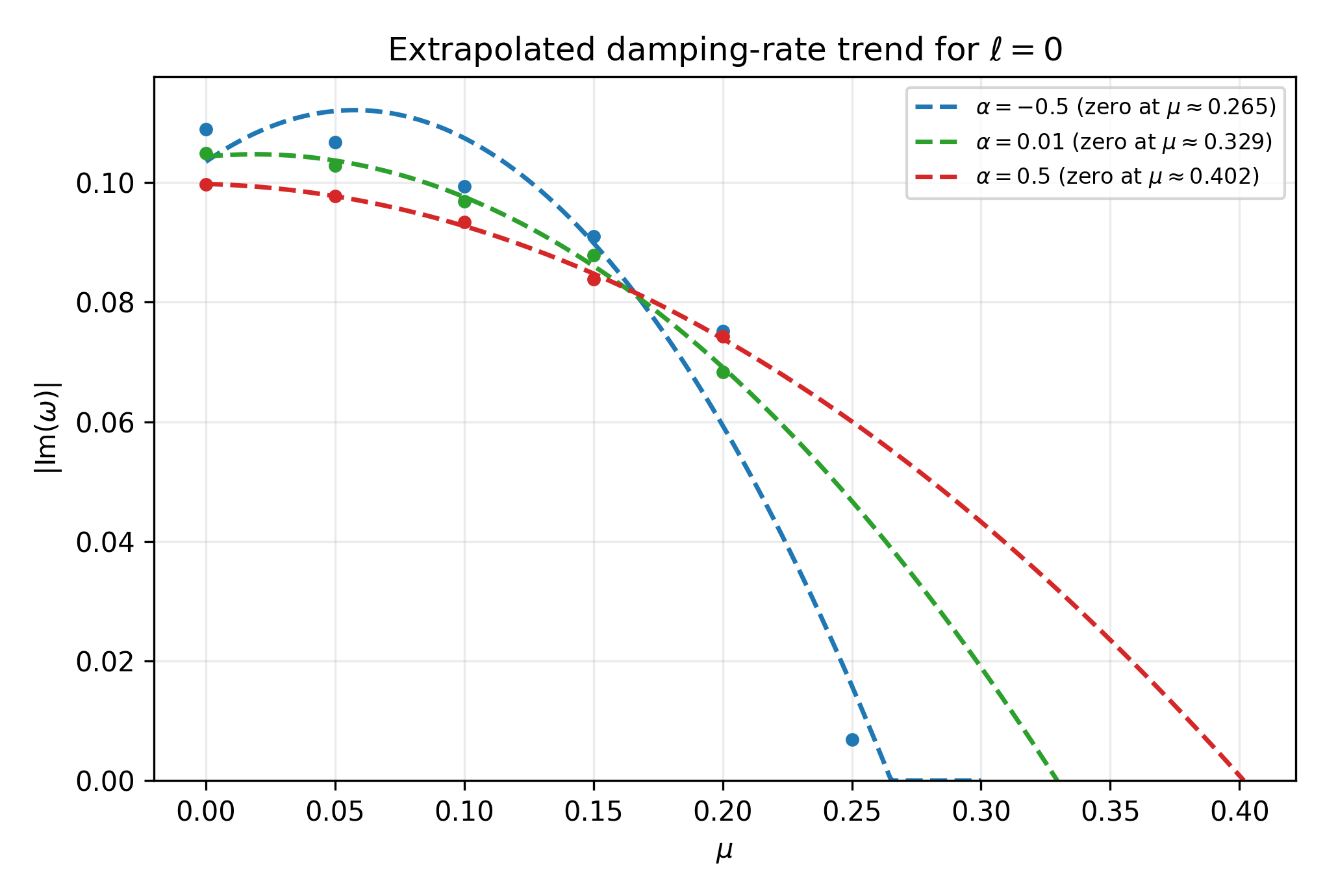}
\includegraphics[width=\columnwidth]{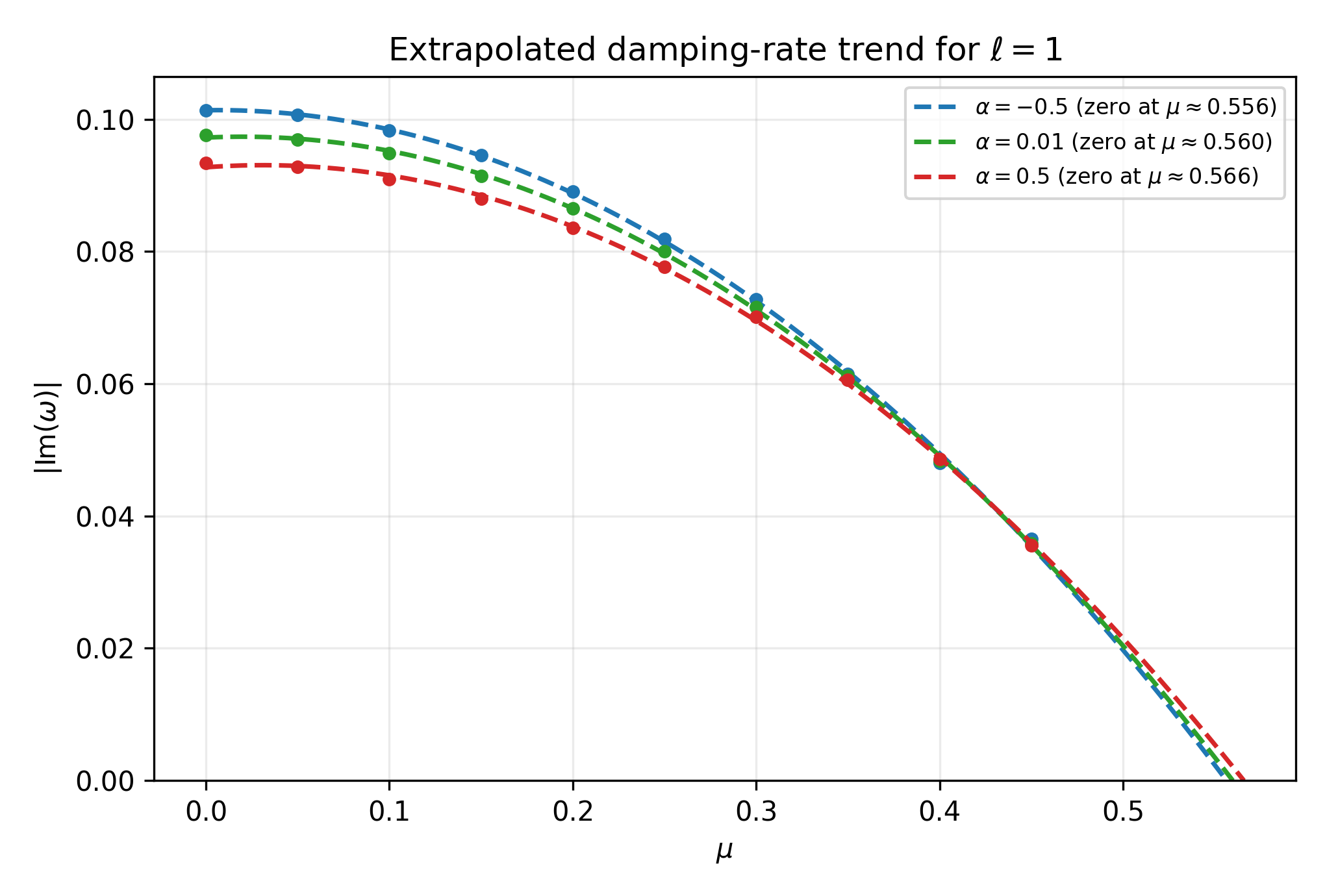}
\includegraphics[width=\columnwidth]{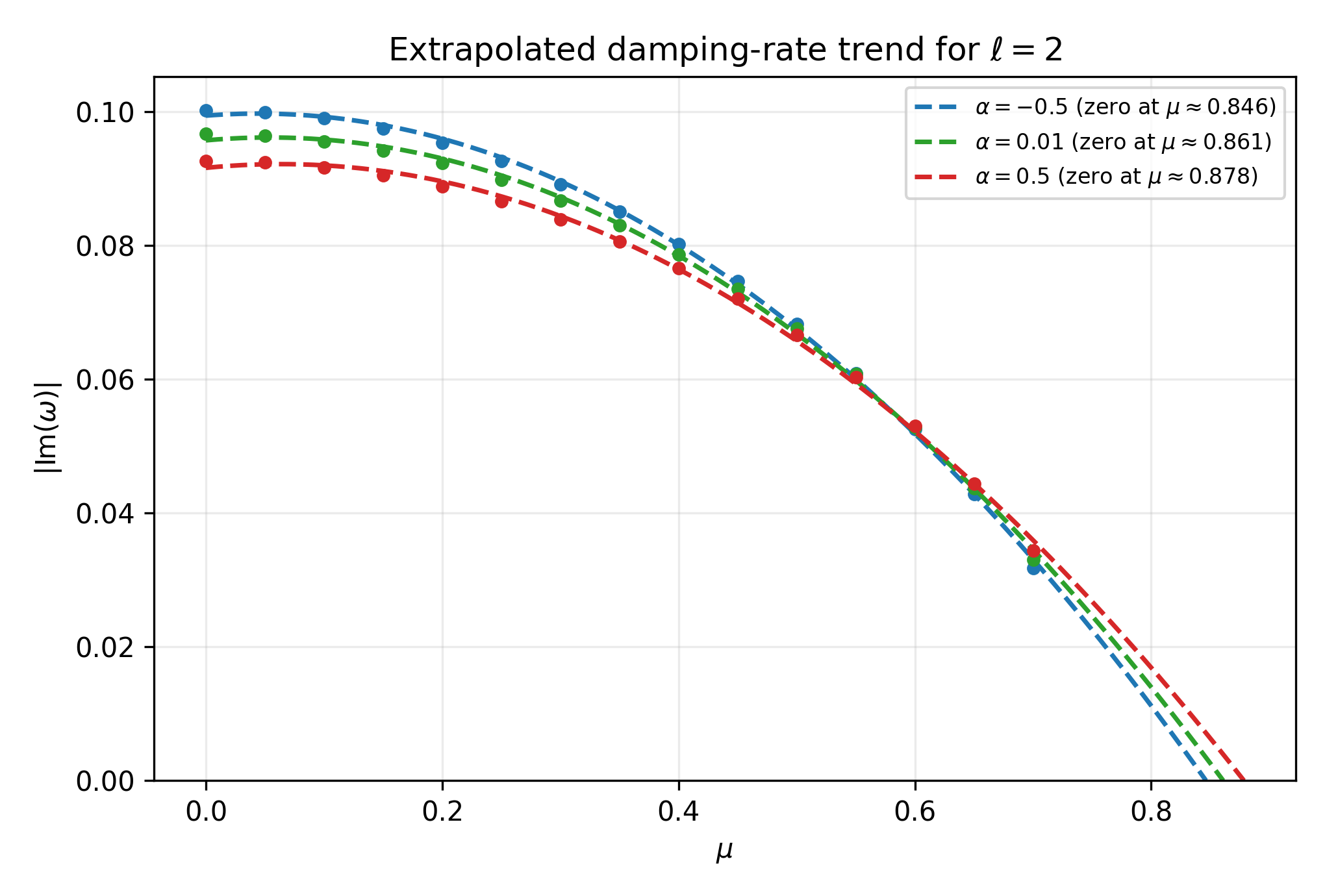}
\caption{Absolute damping rate $|\mathrm{Im}(\omega)|$ versus scalar mass $\mu$ ($M=1$) for fixed multipole $\ell=0,1,2$ and three Gauss--Bonnet couplings $\alpha=-0.5,\,0.01,\,0.5$ (WKB16 data points). Dashed curves show quadratic extrapolations in $\mu$ to the formal zero-damping limit $|\mathrm{Im}(\omega)|\to 0$, illustrating the quasi-resonant trend of progressively longer-lived modes at larger $\mu$. We can see that for $\ell=0$ and $\alpha=-0.5$ the quadratic fit is inaccurate, yet the damping rate evidently approaches zero in this case as well.}
\label{fig:damping_extrap}
\end{figure}

\begin{table}
\textbf{(a) $\alpha=-0.5$, $\ell=0$}\\
\resizebox{\columnwidth}{!}{%
\begin{tabular}{c c c c}
\hline
$\mu$ & WKB16 ($\tilde{m}=8$) & WKB14 ($\tilde{m}=7$) & difference \\
\hline
$0$ & $0.106274-0.108946 i$ & $0.106263-0.108957 i$ & $0.0101\%$\\
$0.05$ & $0.106675-0.106711 i$ & $0.106645-0.106741 i$ & $0.0285\%$\\
$0.1$ & $0.109318-0.099319 i$ & $0.109349-0.099337 i$ & $0.0238\%$\\
$0.15$ & $0.112929-0.091000 i$ & $0.109865-0.090769 i$ & $2.12\%$\\
$0.2$ & $0.104252-0.075144 i$ & $0.113409-0.074489 i$ & $7.14\%$\\
$0.25$ & $0.199834-0.006875 i$ & $0.202089-0.005285 i$ & $1.38\%$\\
\hline
\end{tabular}}

\vspace{0.1cm}
\textbf{(b) $\alpha=-0.5$, $\ell=1$}\\
\resizebox{\columnwidth}{!}{%
\begin{tabular}{c c c c}
\hline
$\mu$ & WKB16 ($\tilde{m}=8$) & WKB14 ($\tilde{m}=7$) & difference \\
\hline
$0$ & $0.287106-0.101370 i$ & $0.287102-0.101370 i$ & $0.0015\%$\\
$0.05$ & $0.288265-0.100620 i$ & $0.288260-0.100620 i$ & $0.0016\%$\\
$0.1$ & $0.291749-0.098358 i$ & $0.291744-0.098358 i$ & $0.0017\%$\\
$0.15$ & $0.297580-0.094545 i$ & $0.297572-0.094544 i$ & $0.0026\%$\\
$0.2$ & $0.305781-0.089097 i$ & $0.305779-0.089098 i$ & $0.0007\%$\\
$0.25$ & $0.316397-0.081904 i$ & $0.316395-0.081905 i$ & $0.0009\%$\\
$0.3$ & $0.329432-0.072781 i$ & $0.329439-0.072795 i$ & $0.0044\%$\\
$0.35$ & $0.344822-0.061500 i$ & $0.344835-0.061431 i$ & $0.0201\%$\\
$0.4$ & $0.362243-0.048006 i$ & $0.362189-0.048089 i$ & $0.0271\%$\\
$0.45$ & $0.385758-0.036511 i$ & $0.388664-0.036276 i$ & $0.752\%$\\
\hline
\end{tabular}}

\vspace{0.1cm}
\textbf{(c) $\alpha=-0.5$, $\ell=2$}\\
\resizebox{\columnwidth}{!}{%
\begin{tabular}{c c c c}
\hline
$\mu$ & WKB16 ($\tilde{m}=8$) & WKB14 ($\tilde{m}=7$) & difference \\
\hline
$0$ & $0.474886-0.100258 i$ & $0.474886-0.100258 i$ & $0.00003\%$\\
$0.05$ & $0.475712-0.099955 i$ & $0.475712-0.099955 i$ & $0.00002\%$\\
$0.1$ & $0.478194-0.099044 i$ & $0.478194-0.099044 i$ & $0\%$\\
$0.15$ & $0.482338-0.097521 i$ & $0.482338-0.097521 i$ & $0\%$\\
$0.2$ & $0.488157-0.095377 i$ & $0.488157-0.095377 i$ & $0\%$\\
$0.25$ & $0.495670-0.092601 i$ & $0.495670-0.092601 i$ & $0\%$\\
$0.3$ & $0.504901-0.089174 i$ & $0.504901-0.089174 i$ & $0\%$\\
$0.35$ & $0.515880-0.085072 i$ & $0.515880-0.085072 i$ & $0.00001\%$\\
$0.4$ & $0.528644-0.080258 i$ & $0.528644-0.080258 i$ & $0.00003\%$\\
$0.45$ & $0.543237-0.074683 i$ & $0.543237-0.074683 i$ & $0.00003\%$\\
$0.5$ & $0.559709-0.068276 i$ & $0.559709-0.068276 i$ & $0.00002\%$\\
$0.55$ & $0.578104-0.060934 i$ & $0.578104-0.060935 i$ & $0\%$\\
$0.6$ & $0.598456-0.052531 i$ & $0.598460-0.052535 i$ & $0.00077\%$\\
$0.65$ & $0.620752-0.042853 i$ & $0.620751-0.042866 i$ & $0.00210\%$\\
$0.7$ & $0.644853-0.031756 i$ & $0.644788-0.031793 i$ & $0.0116\%$\\
\hline
\end{tabular}}
\caption{Fundamental scalar QNMs for $\alpha=-0.5$ and $M=1$, computed with WKB-Pad\'e approximants $\tilde m=8$ (WKB16, i.e., 16th-order WKB) and $\tilde m=7$ (WKB14, i.e., 14th-order WKB). Panels (a)--(c) correspond to $\ell=0,1,2$. The last column reports the relative difference between the two approximants.}
\end{table}

The tabulated frequencies show a clear and systematic mass dependence. For each fixed $(\alpha,\ell)$, increasing $\mu$ generally reduces the damping rate $|\mathrm{Im}(\omega)|$, while the oscillation frequency $\mathrm{Re}(\omega)$ varies more moderately. Physically, this behavior is consistent with increasingly long-lived trapped oscillations of the massive field. In the large-$\mu$ regime, the spectrum approaches the quasi-resonant limit in which $\mathrm{Im}(\omega)\to 0^-$ and the mode lifetime grows strongly.

The comparison between WKB16 and WKB14 provides an internal accuracy diagnostic. In most entries, the relative discrepancy remains very small (typically much less than $1\%$), indicating robust convergence of the high-order WKB-Pad\'e scheme in the parameter region where a single-barrier peak is well defined. Larger deviations appear close to the edges of the allowed mass range, where the potential peak becomes shallow, and WKB assumptions are less favorable. These points should therefore be interpreted with additional caution.

The extrapolation plots in Fig.~\ref{fig:damping_extrap} summarize this trend across multipoles. Although the extrapolated zero-damping points are model-dependent and should not be viewed as exact thresholds, they provide a useful phenomenological estimate of where quasi-resonant behavior sets in for each $(\alpha,\ell)$ branch. The persistence of the trend for $\ell=0,1,2$ supports the conclusion that the long-lived-mode regime is a generic feature of the massive scalar sector in the considered 4D-EGB background.

Overall, the data indicate that the dominant physical effect of increasing field mass is suppression of damping rather than a dramatic shift of oscillation frequency. Combined with the small WKB16--WKB14 spread over most of the scan, this supports the reliability of the present frequency-domain results as a baseline for subsequent time-domain cross-checks and for future parameter-inference applications.

For each multipole $\ell$, the radial equation
\begin{equation}
\frac{d^2\psi}{dr_*^2}+\left[\omega^2-V_\ell(r)\right]\psi=0,
\end{equation}
is solved with quasinormal boundary conditions,
\begin{equation}
\psi\propto e^{\pm i\omega r_*},\quad r_*\to\pm\infty,
\end{equation}
which select a discrete complex spectrum $\omega_{\ell n}=\omega_R-i\omega_I$.

In this work, the QNM frequencies are extracted with higher-order WKB around the potential peak and cross-checked by time-domain fitting, as described above.

\begin{figure}
\includegraphics[width=\columnwidth]{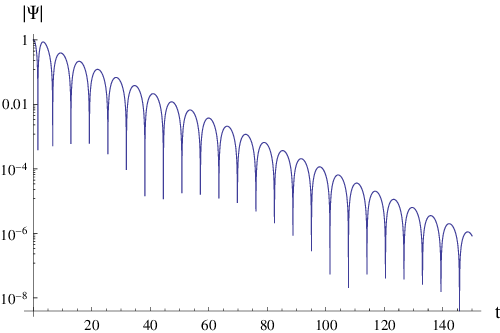}
\caption{Time-domain ringdown profile for the massive scalar perturbation with $(\ell,\mu,\alpha)=(2,0.1,0.5)$ (with $M=1$). The oscillatory tail is fitted by the Prony method, yielding $\omega_{\rm Prony}=0.496502-0.0917142\,i$. For the same parameter set, the frequency-domain WKB16 value from the QNM table is $\omega_{\rm WKB}=0.496501-0.091715\,i$. Taking the time-domain/Prony result as the reference benchmark, the relative complex error of WKB is $|\omega_{\rm WKB}-\omega_{\rm Prony}|/|\omega_{\rm Prony}|\times100\%\approx2.54\times10^{-4}\%$, confirming very good agreement while indicating that the direct time-domain extraction is the more accurate estimate.}
\label{fig:td_profile_l2_mu01_a05}
\end{figure}

The values of QNMs obtained by the WKB method can also be checked by the time-domain integration. An example of the time-domain profile is given in Fig. \ref{fig:td_profile_l2_mu01_a05}, where one can see that the WKB data is in excellent agreement with the time-domain results. 

\section{Grey-body factors and absorption cross-section} \label{sec7}

For GBFs, we consider a scattering setup with asymptotics
\begin{equation}
\psi\to
\begin{cases}
T_\ell(\omega)e^{-i\omega r_*}, & r_*\to-\infty,\\
 e^{-i\omega r_*}+R_\ell(\omega)e^{+i\omega r_*}, & r_*\to+\infty,
\end{cases}
\end{equation}
where $R_\ell$ and $T_\ell$ are reflection and transmission amplitudes. Flux conservation gives
\begin{equation}
|R_\ell|^2+|T_\ell|^2=1.
\end{equation}
The partial GBF is
\begin{equation}
\Gamma_\ell(\Omega)=|T_\ell|^2\Bigg|_{\omega=\Omega},
\end{equation}
where we have introduced a real frequency $\Omega$, which should not be confused with the complex QNMs, which we denote by $\omega$.

Within the WKB approach near the barrier maximum $V_0$, one can write
\begin{equation}
\Gamma_\ell(\Omega)=\left(1+e^{2S_\ell(\Omega)}\right)^{-1},
\end{equation}
with
\begin{equation}
S_\ell(\Omega)=\pi\,\frac{V_0-\Omega^2}{\sqrt{-2V_0''}}+\text{(higher-order WKB corrections)}.
\end{equation}
For $\Omega^2\gg V_0$, one has $\Gamma_\ell\to1$, while for $\Omega^2\ll V_0$, $\Gamma_\ell$ is exponentially suppressed. The WKB method has been used for finding GBFs in \cite{Konoplya:2023ahd,Dubinsky:2025ypj,Malik:2025qnr,Malik:2025erb,Konoplya:2009hv,Lutfuoglu:2025kqp,Lutfuoglu:2025eik,Dubinsky:2025wns}.

\begin{figure*}[t]
\centering
\includegraphics[width=0.32\textwidth]{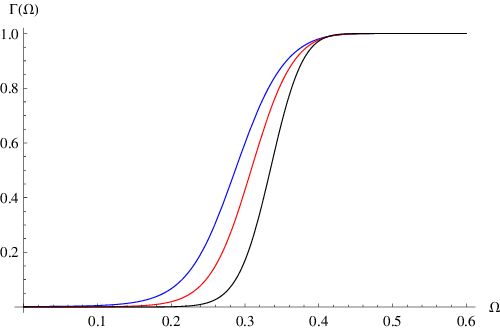}\hfill
\includegraphics[width=0.32\textwidth]{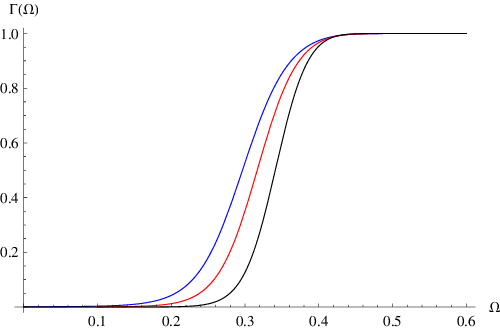}\hfill
\includegraphics[width=0.32\textwidth]{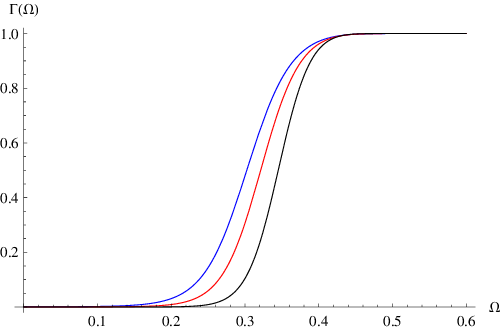}
\caption{Grey-body factors $\Gamma_{\ell=1}(\Omega)$ for the 4D-EGB black hole ($M=1$) at three Gauss--Bonnet couplings: (a) $\alpha=-1$ (left), (b) $\alpha=0.01$ (center), and (c) $\alpha=0.5$ (right). In each panel, the curves correspond to scalar masses $\mu=0$ (blue), $\mu=0.2$ (red), and $\mu=0.3$ (black). The comparison shows how both coupling and field mass shape the transmission probability in the $\ell=1$ channel.}
\label{fig:gbf_l1_combined}
\end{figure*}

As shown in Fig.~\ref{fig:gbf_l1_combined}, increasing the scalar mass generally shifts the transmission rise to higher frequencies, while changing $\alpha$ modifies the detailed slope and onset of the grey-body window.

\begin{figure*}[t]
\centering
\includegraphics[width=0.32\textwidth]{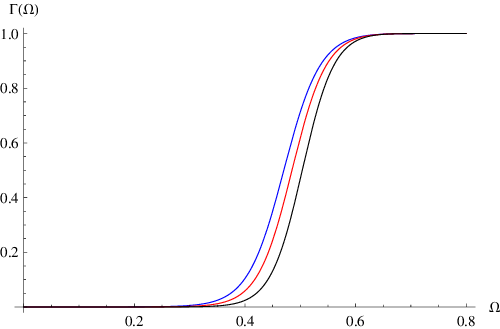}\hfill
\includegraphics[width=0.32\textwidth]{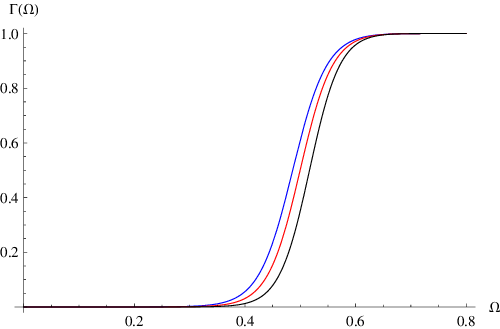}\hfill
\includegraphics[width=0.32\textwidth]{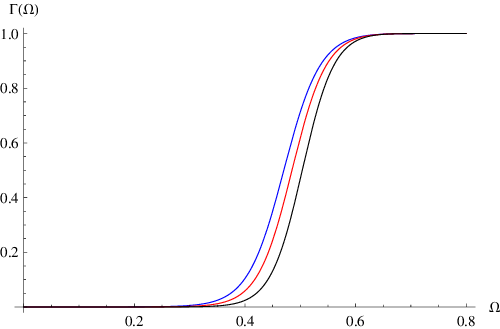}
\caption{Grey-body factors $\Gamma_{\ell=2}(\Omega)$ for the 4D-EGB black hole ($M=1$) at three Gauss--Bonnet couplings: (a) $\alpha=-1$ (left), (b) $\alpha=0.01$ (center), and (c) $\alpha=0.5$ (right). In each panel, the curves correspond to scalar masses $\mu=0$ (blue), $\mu=0.2$ (red), and $\mu=0.3$ (black). The combined view shows the expected low-frequency suppression for $\ell=2$ and its dependence on the Gauss--Bonnet coupling.}
\label{fig:gbf_l2_combined}
\end{figure*}

As shown in Fig.~\ref{fig:gbf_l2_combined}, the $\ell=2$ transmission onset is shifted to higher frequencies compared to $\ell=1$, while varying $\alpha$ changes the detailed shape of the grey-body window.

This behavior follows directly from the barrier interpretation of scattering on the effective potential. A higher and/or wider potential barrier implies a larger WKB action $S_\ell$, hence a smaller transmission probability $\Gamma_\ell=(1+e^{2S_\ell})^{-1}$. Consequently, parameter choices that raise the barrier suppress GBFs at fixed frequency and push the transition to $\Gamma_\ell\sim 1$ to higher $\omega$. In particular, increasing the multipole number $\ell$ strengthens the centrifugal part of $V_\ell$, which is why the $\ell=2$ curves are more suppressed at low frequency than the $\ell=1$ curves. Likewise, increasing field mass $\mu$ lifts the asymptotic level and effectively makes tunneling less efficient in the relevant frequency range, yielding smaller GBFs until sufficiently high frequencies are reached. By contrast, within the considered stable window, changing $\alpha$ produces comparatively modest deformations of the barrier, consistent with the milder $\alpha$-dependence seen in the GBF panels.

For a minimally coupled scalar field, the total absorption cross-section is obtained by summing partial waves:
\begin{equation}
\sigma_{\mathrm{abs}}(\Omega)=\sum_{\ell=0}^{\infty}\sigma_\ell(\Omega),\quad \sigma_\ell(\Omega)=\frac{\pi}{\Omega^2}(2\ell+1)\,\Gamma_\ell(\Omega).
\end{equation}
Hence,
\begin{equation}
\sigma_{\mathrm{abs}}(\Omega)=\frac{\pi}{\Omega^2}\sum_{\ell=0}^{\infty}(2\ell+1)|T_\ell(\Omega)|^2.
\end{equation}
In numerical calculations, the sum is truncated at $\ell=\ell_{\max}$ once convergence is reached.

\begin{figure*}[t]
\centering
\includegraphics[width=0.48\textwidth]{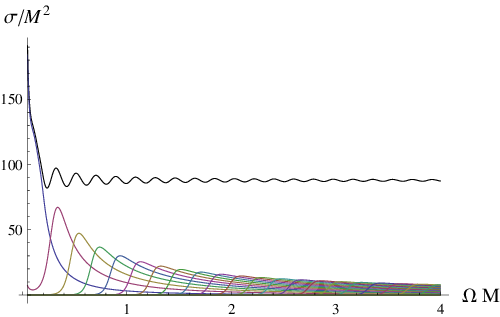}\hfill
\includegraphics[width=0.48\textwidth]{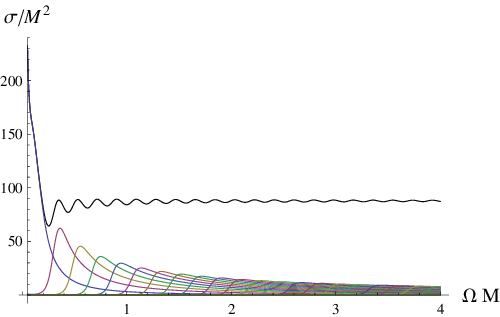}
\caption{Partial and total absorption cross-sections for the 4D-EGB black hole at fixed Gauss--Bonnet coupling $\alpha=-0.5$: panel (a) corresponds to $\mu=0$, and panel (b) to $\mu=0.2$. The two-panel comparison shows how the field mass modifies both the partial-wave structure and the total absorption profile.}
\label{fig:cross_sections_mu_compare}
\end{figure*}

The absorption plots in Fig.~\ref{fig:cross_sections_mu_compare} show a clear mass dependence. For $\mu=0.2$, both the partial contributions and the total absorption are suppressed in the low- and intermediate-frequency region compared to the $\mu=0$ case. This is consistent with the effective-potential picture: increasing $\mu$ raises the asymptotic level of the potential and increases the effective barrier to transmission, so fewer modes penetrate to the horizon at fixed $\omega$. As $\omega$ grows, the barrier becomes progressively less restrictive and the difference between the two mass cases diminishes, recovering the expected high-frequency approach where transmission is less sensitive to $\mu$.

In the partial-wave decomposition, the same trend appears mode by mode: higher $\mu$ shifts the onset of significant contribution from each multipole to larger frequencies, which in turn delays the buildup of the total absorption cross-section.

In the high--frequency (eikonal) regime, the scattering problem simplifies considerably, allowing for an analytic representation of the transmission coefficient in terms of fundamental QNMs of the black hole. In this limit, the GBF takes the form \cite{Konoplya:2024vuj,Konoplya:2024lir}
\begin{equation}\label{transmission-eikonal}
\Gamma_{\ell}(\Omega)=\left(1+\exp\left[2\pi\,\frac{\Omega^2-\re{\omega_0}^2}{4\re{\omega_0}\im{\omega_0}}\right]\right)^{-1}+\Order{\frac{1}{\ell}},
\end{equation}
where $\omega_0$ denotes the least damped quasinormal frequency.

The above expression reflects the fact that, at large $\ell$, both wave scattering and quasinormal ringing are governed by the same local properties of the effective potential in the vicinity of its maximum. The corrections terms, which are written explicitly in \cite{Konoplya:2024lir,Konoplya:2024vuj}, make the correspondence more accurate even at small $\ell$. This correspondence was tested and used in numerous publications \cite{Lutfuoglu:2025blw,Malik:2024cgb,Han:2025cal,Skvortsova:2024msa,Malik:2025dxn,Lutfuoglu:2025hjy,Dubinsky:2025nxv,Bolokhov:2024otn,Lutfuoglu:2025ldc,Lutfuoglu:2025ohb}.

Beyond providing a qualitative interpretation, this connection offers a practical approximation scheme, whose accuracy improves with increasing multipole number. A number of recent analyses have confirmed that the agreement between GBFs reconstructed in this way and those obtained from direct scattering calculations is remarkably good \cite{Lutfuoglu:2025mqa,Bolokhov:2025lnt,Malik:2025czt}.

\section{Conclusions} \label{sec8}

We have developed and applied a consistent framework to study quasinormal ringing and scattering observables of a massive scalar field in the four-dimensional EGB black-hole background. The analysis combines (i) high-order WKB calculations for complex quasinormal frequencies, (ii) a complementary time-domain characteristic-evolution strategy for cross-validation, and (iii) grey-body-factor/absorption diagnostics tied to the same effective potential.

The results show clear and robust mass-driven trends. For fixed $(\alpha,\ell)$, increasing $\mu$ generally reduces damping (smaller $|\im{\omega}|$), indicating longer-lived modes and approach to quasi-resonant behavior. The WKB16--WKB14 agreement is very good for most benchmark points, with larger deviations mainly near edge regions where the barrier is shallow, and WKB assumptions are less favorable.

The scattering sector is consistent with the potential-barrier interpretation. Parameter changes that increase the effective barrier suppress transmission, producing smaller GBFs at fixed frequency and shifting the transition to efficient absorption to higher $\Omega$. This is visible both in the multipole dependence ($\ell=2$ more suppressed than $\ell=1$ at low frequencies) and in the mass dependence of partial/total absorption (the $\mu=0.2$ case is more suppressed than $\mu=0$ until sufficiently high frequencies).

A central methodological point is the stability-constrained parameter choice. Because sufficiently large positive coupling is known to trigger eikonal instability in the gravitational sector, restricting to a conservative stable interval in $\alpha$ is essential to avoid mixing scalar-spectrum trends with backgrounds expected to be dynamically unstable.

Natural next steps include: (a) completing full time-domain extraction and one-to-one comparison against WKB values across the scan, (b) extending the survey to higher overtones and larger multipoles, (c) refining quasi-resonant threshold mapping in $(\alpha,\mu)$ space, and (d) generalizing the same pipeline to broader higher-curvature frameworks, especially four-dimensional Einstein--Lovelock models beyond the Gauss--Bonnet term. Such extensions can test how additional Lovelock corrections reshape both QNM spectra and absorption observables, and whether analogous stability bounds control the phenomenology.

Overall, the present study provides a practical baseline for massive-field spectroscopy in higher-curvature black-hole spacetimes and supports future phenomenological applications in next-generation ringdown and black-hole-scattering analyses.

\begin{acknowledgments}
B. C. L. is grateful to the Excellence project FoS UHK 2205/2025-2026 for the financial support.
\end{acknowledgments}

\bibliography{references}
\end{document}